\documentclass[prl,twocolumn,aps,amssymb,footinbib]{revtex4}
\usepackage{amssymb}
\usepackage{graphicx}
\usepackage{amsmath}
\usepackage{times}
\usepackage{color}
\usepackage{subfigure}
\usepackage{setspace}
\usepackage{bm}
\usepackage{multirow}

\newcommand {\beq} {\begin{equation}}
\newcommand {\eeq} {\end{equation}}
\newcommand {\bqa} {\begin{eqnarray}}
\newcommand {\eqa} {\end{eqnarray}}

\def \be{\begin{equation}}
\def \ee{\end{equation}}
\def \ba{\begin{array}}
\def \ea{\end{array}}
\def \bea{\begin{eqnarray}}
\def \eea{\end{eqnarray}}

\def \l{\left}

\def \rr{\right}

\begin{document}

\begin{abstract}
  We consider a system composed of a stack of weakly Josephson coupled
  superfluid layers with c-axis disorder in the form of random
  superfluid stiffnesses and vortex fugacities in each layer as well
  as random inter-layer coupling strengths. In the absence of disorder
  this system has a 3D XY type superfluid-normal phase transition as a
  function of temperature. We develop a functional renormalization
  group to treat the effects of disorder, and demonstrate that the
  disorder results in the smearing of the superfluid normal phase
  transition via the formation of a Griffiths phase. Remarkably, in
  the Griffiths phase, the emergent power-law distribution of the
  inter-layer couplings gives rise to sliding Griffiths superfluid,
  with an anisotropic critical current, and with a finite stiffness in a-b direction along the layers, and a
  vanishing stiffness perpendicular to it.  
\end{abstract}

\title{Finding the elusive sliding phase in superfluid-normal phase
  transition smeared by c-axis disorder}

\author{David Pekker$^1$, Gil Refael$^2$, Eugene Demler$^1$}  
 \affiliation{$^1$ Physics Department,
Harvard University, 17 Oxford st., Cambridge, MA 02138\\
$^2$ Physics Department,
California Institute of Technology, MC 114-36, 1200 E. California
Blvd., Pasadena, CA 91125}
\maketitle

The interplay of disorder and broken symmetry remains a challenging
and relevant problem for correlated quantum systems.  The effects of
disorder in one dimensions, where the effects of quantum fluctuations
are enhanced, is most dramatic, giving rise to Anderson
localization~\cite{Anderson}, Dyson singularities, and random singlet
phases~\cite{Fisher1994_95}. Recent studies, both experimental and
theoretical, concentrated on the superfluid-insulator transition of
Bosonic chains~\cite{localization2008}, and strongly argued that
disorder alters the universality of that
transition~\cite{AKPR1,AKPR2}.  While uncorrelated disorder in higher
dimensions has a lessened effect, we must raise the question: how does
correlated disorder, which only varies in a subset of directions,
affects thermal and quantum phase transitions in higher dimensions?

In this work, we study this question by concentrating on the
superfluid insulator transition in 3D Bose gases, that is split into a
series of pancake clouds by a 1D optical lattice with disorder which
varies only along the lattice direction, but not parallel to the
clouds. While this question is of much theoretical interest, and is
now also of experimental relevance as we outline below, it was not
addressed so far. The effects of the disorder could be as mundane as
just shifting the transition point, or as important as resulting in a
new universality class of the transition or obliterating it
altogether. Indeed, we shall show that the interplay between disorder
along the c-axis and the a-b plane Berezinskii-Kosterlitz-Thouless
(BKT) physics~\cite{Berezinskii, KosterlitzThouless1973}, smears the
transition giving rise to an intermediate Griffiths
phase~\cite{Vojta2006, Vojta2008} that occupies a wide region of the
phase diagram. Furthermore, in a subphase within this Griiffith phase,
the superfluid becomes split into an array of 2D puddles that have no
phase coherence along the c-axis, thus realizing the illusive sliding
phase paradigm~\cite{Toner1999}, supporting superflow only in the a-
and b- but not c-directions.
\begin{figure}
\includegraphics[width=8.5cm]{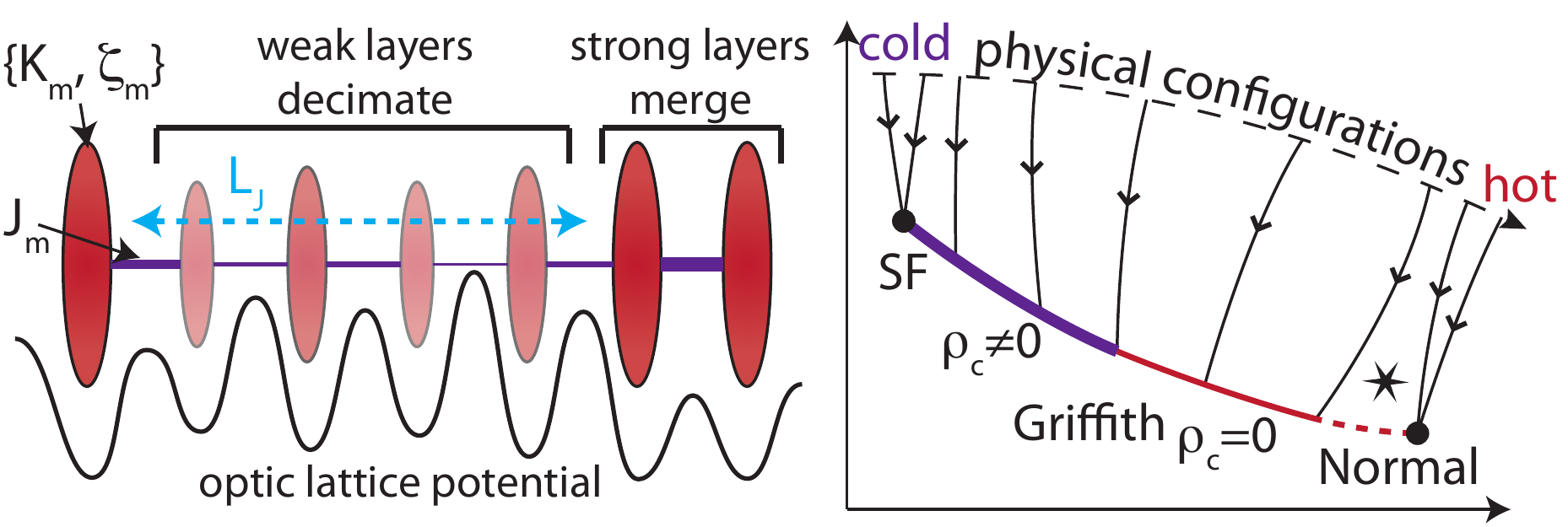}
\caption{Left: Schematic diagram of the model: red ovals (purple
  bars) represent the superfluid layers (Josephson couplings) with
  size inversely proportional to vortex fugacity $\zeta_m$ (directly 
  proportional to Josephson coupling $J_m$).  
  The effects of the real space RG is to merge strong layers, and
  decimate weak layers. Emergent length-scale $L_J$
  corresponds to the typical separation between strong layers. \\
  Right: Schematic diagram of the RG flows showing the superfluid (SF)
  and normal fixed points along with the Griffiths fixed line. The
  black dashed line represents physical configurations, with points to
  the right corresponding to higher temperatures. The Griffiths fixed
  line is split into two segments, corresponding to the regimes with
  finite and zero c-axis superfluid response. The star indicates a
  possible unstable fixed point~\cite{star}.  }
\label{Fig:SchematicFlows}
\end{figure}

The questions we raise are fast becoming important for
experiments. Experiments on ultracold atoms observed both the BKT
transition in large 2D ``pancakes'' produced by very deep 1D optical
lattices~\cite{Dalibard2006}, and Anderson localization of Bosons in
1D disordered optical lattices~\cite{localization2008,
  Fallani2008}. The system we study here can be realized by
constructing a stack of large 2D ``pancakes'' using a disordered 1D
optical lattice and tests the effects of disorder near the 2D-3D
crossover~\cite{Affleck1996, Kasevich}.

The model which we analyze and describe below consists of a set of
coupled 2D superfluid layers. Each layer has a superfluid stiffness
$K_m$, vortex fugacity (akin to vortex density per coherence length)
$\zeta_m$, and Josephson coupling (to the next layer) $J_m$. $K_m$,
$\zeta_m$ and $J_m$ are initially random and
uncorrelated~\footnote{Throughout we assume bounded distributions of
  couplings, as otherwise the system would always become
  disconnected.}, see Fig.~\ref{Fig:SchematicFlows}a.  To analyze this
model, we combine a Kosterlitz-Thouless like momentum space
renormalization for the in-plane degrees of
freedom~\cite{KosterlitzThouless1973, Giamarchi2006} with a real-space
RG (see, e.g., \cite{Fisher1994_95, AKPR2}). In the real space-RG
decimation, strongly coupled layers ($J_m\sim 1$) are merged, while
vortex-ridden layers ($\zeta_m\sim 1$) are considered to be essentially
normal and are perturbatively eliminated.

Before plunging into the analysis, let us summarize the phase diagram
we find, see Fig.~\ref{Fig:SchematicFlows}b and
Table~\ref{Table:Phases}. At low temperatures the system forms a 3D
superfluid. As the temperature is raised, a Griffiths phase appears;
in it, the system breaks up into 2D superfluid puddles, each composed
of one or several ``pancakes'', with weak (power law distributed)
inter-puddle tunneling.  As the temperature is increased further, the
c-axis superfluid response disappears altogether, while the system
remains superfluid in the a- and b-directions, realizing a sliding
phase. At yet higher temperatures, the in-plane superfluid response
smoothly vanishes as the system becomes fully normal.

\begin{table}
\begin{tabular}{c|ccccc}
$T$ & phase & $\rho_{ab}$ & $\rho_c$ & $J_{c,ab}$ & $J_{c,c}$ \\
\hline
high & Normal & zero & zero & zero & zero \\
\hline
\multirow{2}{*}{} & Griffith-Sliding & \multirow{2}{*}{finite} & zero & \multirow{2}{*}{finite} & \multirow{2}{*}{zero} \\
& Griffiths & &  finite \\
\hline
low & Superfluid & finite & finite & finite & finite \\
\hline
\end{tabular}
\caption{Phase diagram indicating the properties of the various phases.}
\label{Table:Phases}
\end{table}

The Griffiths phase is perhaps the most surprising aspect of our
results.  At intermediate temperatures, the flow leads to a fixed line
characterized by a stationary power-law distribution of inter-layer
tunnelings, $J_m$, which are $P(J)\sim J^{\nu_J-1}$
(Fig.~\ref{Fig:SchematicFlows}b).  The appearance of these Griffiths
phase power laws are a direct consequence of the disorder.  Most
layers have strong fluctuations, and turn insulating; Neighboring
layer of either side, can still exchange bosons but with a smaller
amplitude, e.g., $J_\text{eff}=J_{m-1}\cdot J_m$ if layer $m$ is
eliminated.  The elimination of all incoherent layers marks a first
epoch in the RG flow, and upon its end the internally coherent layers
are separated from each other by, on average, $L_J$ incoherent layers
Fig.~\ref{Fig:SchematicFlows}a.  $L_J$ determines $\nu_J$: $\nu_J\sim
\log[1/\bar{J}]/L_J$, with $\bar{J}$ the typical initial Josephson
coupling. In the subsequent RG epoch, layers only merge, but $\nu_J$
remains unchanged.

The Griffiths phase can be separated into two regimes. A sliding
regime with $\nu_J$ flowing to $\nu_J<1$ (as indicated in
Fig.~\ref{Fig:SchematicFlows}b), where there is no c-axis stiffness,
and a Griffiths superfluid with a finite c-axis stiffness and
$\nu_J>1$.  Both regimes have a vanishingly small c-axis critical
current. To wit, the critical current of $n$ layers is determined by
the weakest effective tunneling between them. The expectation value
for the longest run of weak layers is ${\cal R}_n\sim
\log_{1/p_\text{weak}} \left[ n (1-p_\text{weak})
\right]$~\cite{Schilling1990} (with $p_\text{weak}$ the probability of a
layer to be normal in the first epoch; $L_J=(1-p_\text{weak})^{-1}\gg
1$). The weakest link is thus $I_c\sim
\l(\frac{n}{L_J}\rr)^{L_J\log \overline{J}}$.

{\it Model --- \/} Let us now describe the model and its analysis more precisely, before
discussing more of its consequences and experimental implications. 
Following Ref.~\cite{Giamarchi2006}, our model consists
of a set of coupled $1+1$ dimensional (Euclidean) sine-Gordon
models with partition function $Z=\text{Tr} \, \exp-\sum_m \left(S_{sG;m}+S_{J;m,m+1}\right)$ where
\begin{align}
S_{sG;m}&=\int dy dx \big[ K_m (\partial_x \theta_m)^2+\frac{1}{K_m}
(\partial_x \phi_m)^2 \nonumber \\
&\quad\quad-2 i (\partial_x\phi_m)(\partial_y \theta_m)+\zeta_m \cos(2 \phi_m)\big],\\
S_{J;m,m+1}&=\int dy dx
, J_m \cos(\theta_m-\theta_{m+1}).
\end{align}
$S_{sG;m}$ is the sine-Gordon action that describes the density waves
and vortices in the $m$-th layer; $S_{J;m,m+1}$ is the $m$ to $m+1$
tunneling; $\theta_m(x,y)$ and $\phi_m(x,y)$ are the superfluid
order-parameter phase variable and its conjugate, respectively
($[\theta_m(r),\partial_x \phi_{m'}(r')]=i \pi
\delta(r-r')\delta_{m,m'}$). We define $J_m={\cal J}_m/T$, $K_m={\cal
  K}_m/T$, and $\zeta_m \sim \exp(-E_{\text{core},m}/T)$ as the
reduced Josephson coupling, superfluid stiffness, and vortex fugacity
at temperature $T$, where $E_{\text{core},m}$ is the vortex core
energy. We note that to define the sine-Gordon model we must specify
the short-distance cut-off scale. We choose a single cut-off $a$ for
all layers, and work in the units in which $a=1$.


\begin{figure}
\includegraphics[width=5.5cm]{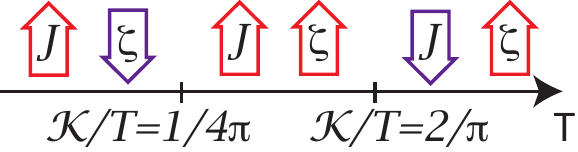}
\caption{Schematic representation of relevance of $J$'s and $\zeta$'s
  as a function of temperature. }
\label{Fig:SchematicDirections}
\end{figure}

{\it Renormalization Group --- \/} Our analysis relies on a combined
c-axis real space and a-b momentum space RG. 
The momentum space transformation is given by~\footnote{These equations are similar to those of
  Ref.~\cite{Giamarchi2006}, but adapted to take into account un-equal
  coupling constants in neighboring layers.}:
{\small
\begin{align}
\frac{dJ_m}{d\ell}&=\!J_m \!\left[2\!-\!\frac{1}{4\pi K_m}\!-\!\frac{1}{4 \pi K_{m+1}}\right]\!-\!\frac{\pi^2}{2} J_m \!\left(\zeta_m^2\!+\!\zeta_{m+1}^2\right), \label{Eq:msJ}\\
\frac{d\zeta_m}{d\ell}&=\zeta_m \left[2-\pi K_m\right]-\frac{1}{2}\pi^2 \zeta_m (J_m^2+J_{m-1}^2),  \label{Eq:msZeta}\\
\frac{dK_m}{d\ell}&=-2\pi^3 (K_m \zeta_m)^2+\frac{\pi}{2}(J_m^2+J_{m-1}^2).
\label{Eq:msRG}
\end{align}}To lowest order in $\zeta_m$ and $J_m$, there is a range of
superfluid stiffnesses $1/4\pi \lesssim K_m \lesssim 2/\pi$ in which
both the vortex fugacity $\zeta_m$ and the Josephson coupling $J_m$
are relevant. The competition between the two gives rise to the
Griffiths phase. Outside this range the system is either strongly
superfluid (large $K_m$) or strongly insulating (small $K_m$), Fig.~\ref{Fig:SchematicDirections}.

As the in-plane momentum shell RG proceeds, the real-space RG (RSRG)
merges layers where $J_m$ rises to 1, or eliminates layers where
vortex fugacity $\zeta_m$ reaches to 1. When a Josephson coupling
becomes large, $J_i=1$, the relative phase $\Delta
\theta=\theta_{i+1}-\theta_i$ of the two neighboring layers becomes
locked and the two layers merge into a single super-layer having
$K_\text{eff}=K_i+K_{i+1}$ and
$\zeta_\text{eff}=\zeta_i\cdot\zeta_{i+1}$. Similarly, if one of the
vortex fugacities becomes large, $\zeta_i=1$, then the conjugate field
$\phi_i$ in that layer becomes locked and vortices proliferate.  Upon
integrating out the incoherent layer, we find that it suppresses
tunneling across it to $J_\text{eff}=J_{i-1}\cdot J_i$. These RG
rules make it convenient to parametrize $J$ and $\zeta$ in terms of
their logs $j=\log(1/J),\,z=\log(1/\zeta)$ which yields:
\begin{align}
 j_\text{eff}&=j_{i-1}+j_i& z_\text{eff}&=z_{i-1}+z_i.
\label{Eq:rsrg}
\end{align}

Next, instead of a numerical analysis of the RG outlined above (which
we will fully pursue in a separate work ~\cite{LongPaper}), let us use
the RG procedure to derive the approximate flow of the distribution
functions for $K$,\,$j$,\, and $z$, and their universal aspects.
First, note that the RSRG layer merging step leads to strong
correlations of $K$'s and $\zeta$'s. Therefore, alongside the
distribution $P_j$ for $j_m$'s, we use the joint probability
distribution $Q_K^z$ for $z_m$'s and $K_m$'s. The fRG equations
resulting from Eqs. (\ref{Eq:msRG}, \ref{Eq:rsrg}) and the:
\begin{align}
\frac{d P_j}{d\ell}&=I_1 \, \partial_j P_j -\int d K_1 Q_{K_1}^{0} \{ 2-\pi K_1 \}  P_j \nonumber \\
&\!\!\!+\!\int \!\!d K_1 Q_{K_1}^{0} \{ 2\!-\!\pi K_1 \} \int \! dj' P_{j'}P_{j-j'} +I_1 P_0 P_j, \label{Eq:fRGP}\\
\frac{d Q_K^z}{d\ell}&=\!(2\!-\!\pi K) \partial_z Q_K^z \!-\!\!\int\!\! d1\, Q_1\! \left( \!2\!-\!\frac{1}{4 \pi K_1}\!-\!\frac{1}{4 \pi
    K}\!\right) \!Q_K^z\! P_{0}\nonumber \\
&+\int d1\, Q_1 Q_{K-K_1}^{z-z_1} \left( 2-\frac{1}{4 \pi K_1}-\frac{1}{4 \pi
    (K-K_1)}\right)P_{0}\nonumber\\
&+\int dK_1 \, Q_{K_1}^0 \{ 2-\pi K_1 \} Q_K^z, \label{Eq:fRGQ}
\end{align} 
where $I_1=\int d1 \, d2 \, Q_1 Q_2 \left(2-\frac{1}{4 \pi
    K_1}-\frac{1}{4 \pi K_2}\right)$, $\{ g \}$ stands for $g
\Theta(g)$ with $\Theta$ being the step function; $d1$ and $Q_1$ are
shorthand for $dK_1 dz_1$ and $Q_{K_1}^{z_1}$ where it is unambiguous.
Briefly, the first terms of Eqs.~(\ref{Eq:fRGP}) and (\ref{Eq:fRGQ})
correspond to the action of the linear in $J_m$ and $\zeta_m$ terms of
the momentum space RG Eqs.~(\ref{Eq:msJ}) and (\ref{Eq:msZeta}). The
remaining terms correspond to the action of the real space RG, where
we keep in mind the fact that the distributions must be
normalized. The normalization is accomplished by rescaling the
distributions $P_j$ and $Q_K^z$ whenever layers are removed from the
system. The fRG equations must be supplemented by absorbing wall
boundary conditions $Q_K^{0^-}=0$ and $P_{0^-}=0$ which remove the
small $z$'s and $j$'s (large $\zeta$'s and $J$'s) from the
distributions when layers are decimated or merged. In order to compute
physical observables we also keep track of $n(\ell)$, the number of
surviving layers at RG scale $\ell$:
\begin{align}
\frac{dn}{d\ell}=-n\left(I_1 P_0 +\int dK_1\,\{2-\pi K_1\} Q_{K_1}^0 \right). \label{Eq:n}
\end{align}
Note that the structure of the fRG and of the resulting flows are
similar to those in Ref.~\cite{Vojta2008, Rosenow} for the damped
transverse field Ising model~\footnote{There is, however, an important
  difference in that the previous treatment used an energy scale to
  drive the real space RG, while here we use a short-length scale
  cut-off to drive the momentum space RG which naturally results in a
  real space RG.}.

To study the evolution of $P_j(l)$ and $Q_K^z(l)$ under coarse
graining, we numerically integrate Eqs.~(\ref{Eq:fRGP}) and
(\ref{Eq:fRGQ}). To parametrize the initial distributions at
temperature $T$ (and length scale $a$), we choose smooth functions
with the following bounds: $0.04<T J<0.11$, $1.0<T K<1.5$, and
$e^{-1.6 \pi\cdot 1.5 /T} < \zeta <e^{-1.6 \pi\cdot 1.0 /T}$~\footnote{We note
  that the qualitative results are independent of the particular
  choice of distributions as long as they are bounded.}.

\begin{figure*}
\includegraphics[width=12cm]{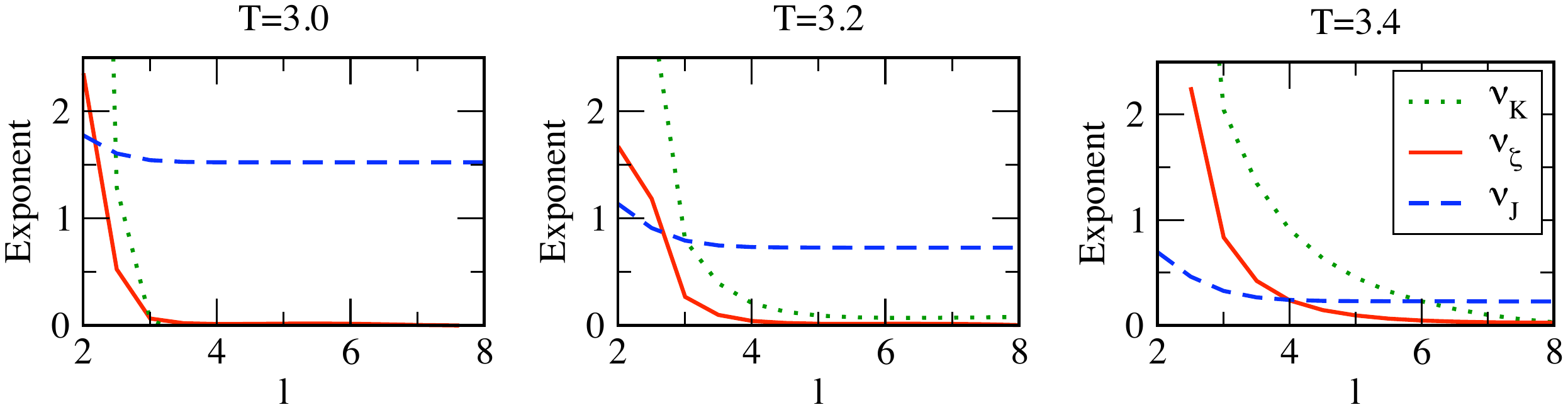}
\hspace{1cm}
\includegraphics[width=4.2cm]{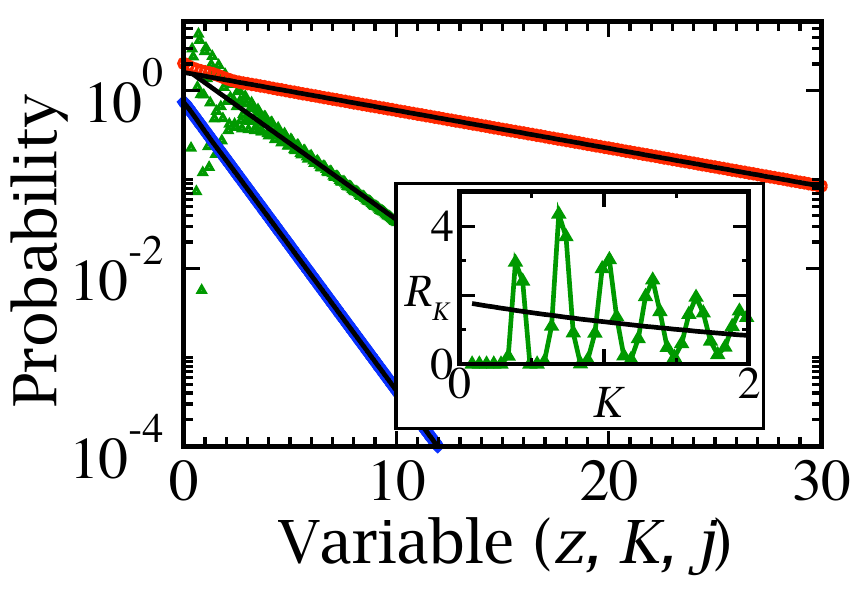}
\caption{Left three panels: flow of exponents $\nu_\zeta$, $\nu_K$,
  and $\nu_J$ under the action of the coarse graining transformation
  for three different initial distributions corresponding to
  temperatures $T=3.0$, $3.2$, and $3.4$.  $l\sim 3$ separates the
  first epoch, in which all three exponents flow from the second epoch
  in which only $\nu_\zeta$ and $\nu_K$ flow. The asymptotic value of
  $\nu_J$ at long length-scales indicates that $T=3.0$ corresponds to a
  Griffiths superfluid, while $T=3.2$ and $3.4$ correspond to Griffith
  insulator. \\
  Right panel: semi-log plot of typical distributions (from top to
  bottom) $Q[z]$, $R[K]$, and $P[j]$, obtained by solving
  Eqs.~(\ref{Eq:fRGP}) and (\ref{Eq:fRGQ}) numerically and the
  corresponding exponential fits (black solid lines) that are used to
  obtain the values of exponents $\nu_\zeta$, $\nu_K$, and $\nu_J$.
  The inset depicts the distribution $R[K]$ on a linear-linear plot,
  the oscillations arise from additions of $K$ values when layers
  merge.  }
\label{Fig:Exp}
\end{figure*}

{\it Results --- \/} A numerical analysis of the flows reveals three
phases: (1) superfluid phase -- all layers merge, (2) Griffiths phase
-- power law distributions $P_J\sim J^{\nu_J-1}$ with a finite
$\nu_J$, (3) insulating phase -- all layers decimate. Phases (1) and
(3) correspond to the usual superfluid and insulating fixed points,
the Griffiths phase, regime (2), corresponds to a new fixed line that
is induced by disorder~\cite{star}.

Within the Griffiths phase, the flow of $P_j$ and $Q_K^z$
distributions occurs in two epochs, as depicted in Fig.~\ref{Fig:Exp}.
In the first epoch both mergers and layer eliminations take place,
which quickly results in the formation of power law distributions with
flow of all three exponents. However, as mergers lead to ever
increasing $K$'s while eliminations do not, the flow eventually
exhausts all weak layers by eliminating them, and only the strongly
superfluid layers remain. In the second epoch, only $J$'s (but not
$\zeta$'s) remain relevant as all the surviving $K$'s exceed $2/\pi$,
so while layers continue to merge, no eliminations occur. As a result,
$\nu_J$ saturates while both $\nu_K$ and $\nu_\zeta$ decay to zero
exponentially in the fRG flow parameter $\nu_K \sim \nu_\zeta \sim
e^{-2 \nu_J \ell}$. We see, therefore, that the Griffiths fixed line
corresponds to a line of fixed $\nu_J$'s with $\nu_K \rightarrow 0$
and $\nu_{\zeta} \rightarrow 0$.

\begin{figure}
\includegraphics[width=6cm]{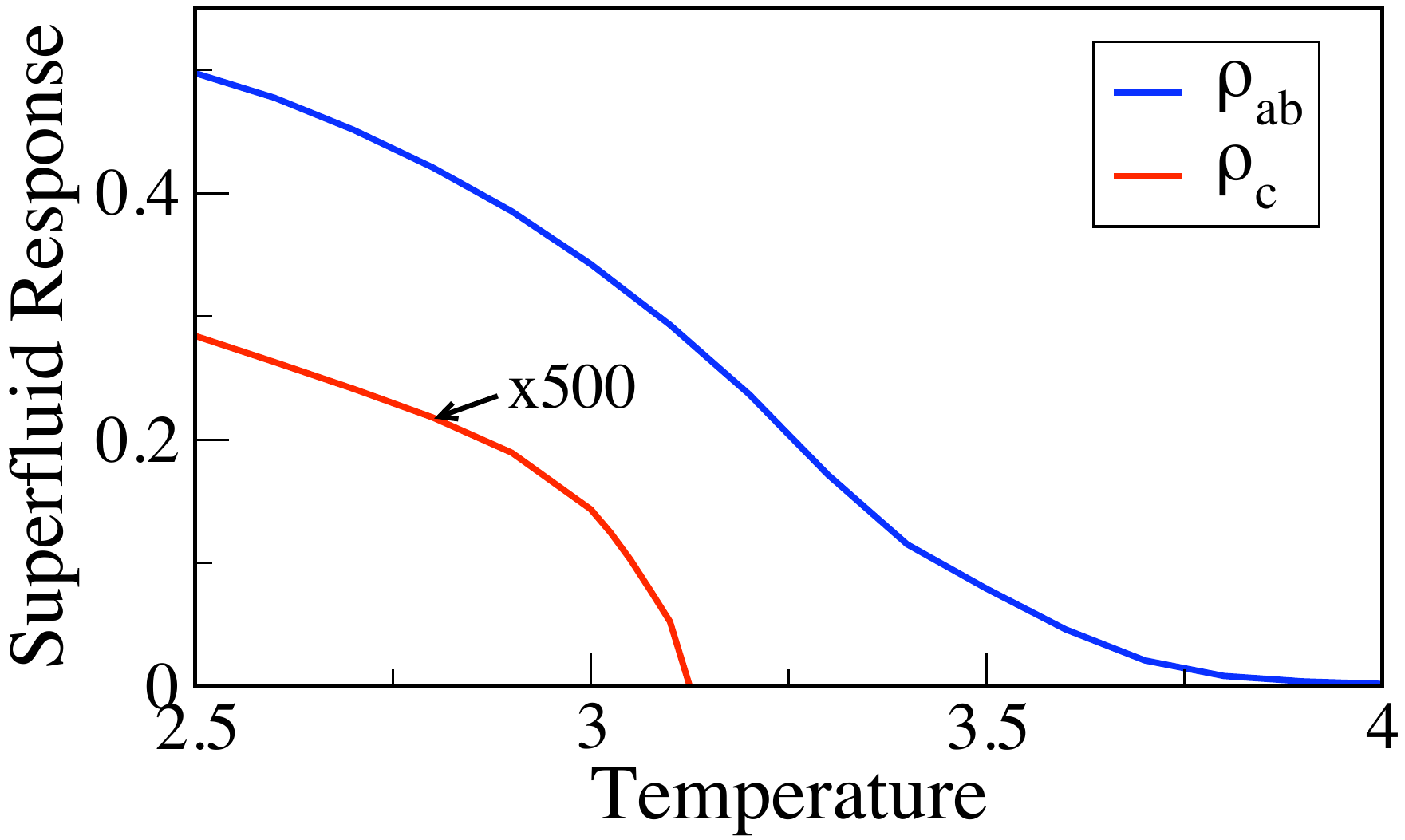}
\caption{The in-plane ($\rho_{ab}$) and out-of-plane ($\rho_c$)
  stiffness as a functions of temperature evaluated using value of RG
  parameter is sufficiently large ($l=10$) so that the flows are
  saturated. While $\rho_{ab}$ undergoes a smeared phase transitions, as
  indicated by the absence of a critical point with power law
  behavior, $\rho_c$ undergoes a continuous phase
  transition at $T \sim 3.1$.}
\label{Fig:Response}
\end{figure}

Both the in-plane $\rho_{ab}$ and the out-of-plane $\rho_c$ superfluid
responses have a very peculiar behavior within the Griffith phase, and
can be used to as probes. The mean values of the superfluid responses
may be obtained from the distributions via
\begin{align}
\rho_{ab} &=n \int dz\, dK K Q_K^z, \\
\rho_c  &=\chi(l)+ e^{-2l}\left(n \int dj e^j P_j\right)^{-1}, \label{Eq:rc}
\end{align} 
where, $n$, which is given by Eq.~(\ref{Eq:n}), is the fraction of
surviving layers and is needed to normalize the response to the scale
of the original system. $\chi(l)$ obeys
\begin{align}
\partial_l \chi(l)&=n P_0 e^{2l} I_1&\chi(0)&=0,
\end{align}
and accounts for stiffness within the superfluid ``puddles''.  As
$\rho_c$ depends on the area, we include the factor of $e^{-2l}$ in
Eq.~\ref{Eq:rc} to account for its renormalization. We plot
$\rho_{ab}$ and $\rho_c$ as a function of $T$ near saturation (at
large value of the RG parameter $\ell$) in
Fig.~\ref{Fig:Response}. The smearing of the phase transition is
reflected in $\rho_{ab}$ decreasing smoothly, as the temperature is
increased, until reaching zero at the end of the Griffiths fixed line
(without following any power law). On the other hand, $\rho_c$
decreases much faster, becoming zero within the Griffith phase at the
point where $\lim_{\ell\rightarrow \infty}\nu_J(\ell)$ becomes smaller
than unity. The disappearance of $\rho_c$ signals the onset of the
elusive sliding subphase of the Griffith phase, see
Table~\ref{Table:Phases}.

{\it Experimental probes --- \/} In our analysis we found that
disorder smears the superfluid-to-normal phase transition. This could be probed
experimentally in the ultra-cold atom setting.  The superfluid
response as well as the critical current could be measured by jolting
the confined gas (e.g. quickly displacing the trap potential) and looking at
the decay of the center of mass oscillations~\cite{McKay2008}.
Alternatively, one could look at correlations by removing the optical
lattice and the trap potentials and allowing the atoms from the
various pancakes to expand and interfere. The key signature of the
Griffiths phase, in interference experiments, is very strong shot noise
which results from the interference of several weakly coupled
superfluid droplets~\cite{LongPaper}. Alternatively, in the mesoscopic
setting, the Griffiths phase could appear in artificially grown
structures composed of alternating layers of superconducting and
insulating films of varying thicknesses. In this setting the anisotropy
superfluid responses and critical currents could be measured directly.

{\it Conclusions --- \/} In this manuscript we investigated the
effects of reduced dimensionality disorder on a
phase transition at a higher dimensionality. We focus on a
model that we believe is relevant to experiments in ultra-cold
gases in optical lattices, and mesoscopic systems such as stacked
superconducting films. Naively, one would expect that, the
strong disorder picture of Ref.~\cite{Fisher1994_95} should be in
effect. However, the classification scheme of Ref.~\cite{Vojta2006}
indicates that the existence of the BKT transition for a single 2D
layer boosts the importance of disorder in our system, resulting in
the stronger effect of the smearing of the phase transition. Using a
functional renormalization group scheme that we develop, we show that
this is indeed the case. Further, we show that in the transition
region the system becomes essentially two dimensional. We find that
the reduction of dimensionality is reflected in the strong anisotropy
of physical observables like critical current and superfluid response.

As we were finalizing this manuscript, we became aware of a
complementary investigation of the random superfluid stack from a
scaling perspective by Vojta and Narayanan, which is consistent with our findings.

{\it Acknowledgements} It is our pleasure to thank Michael Lawler,
Subir Sachdev, and Bryan Clark for insightful discussions. DP and ED
acknowledge support from DARPA, CUA, and NSF Grant
No. DMR-07-05472. GR gratefully acknowledges support from the Packard Foundation,
the Sloan Foundation, and the Cottrell Scholars program.


\begin{thebibliography}{99}
\bibitem{star} We suspect that there may be an additional unstable
  fixed point, indicated by the star in
  Fig.~\ref{Fig:SchematicFlows}b. However, due to the complexity of
  the fRG equations, we were unable to find it or rule it out.
\end{thebibliography}
\end{document}